# Vortex twins and anti-twins supported by multi-ring gain landscapes

Olga V. Borovkova, Yaroslav V. Kartashov, Valery E. Lobanov, Victor A. Vysloukh, and Lluis Torner

*ICFO-Institut de Ciencies Fotoniques, and Universitat Politecnica de Catalunya, Mediterranean Technology Park, 08860 Castelldefels (Barcelona), Spain*
*Corresponding author: Olga.Borovkova@icfo.es*

We address the properties of multi-vortex soliton complexes supported by multi-ring gain landscapes in focusing Kerr nonlinear media with strong two-photon absorption. Stable complexes incorporating two, three, or four vortices featuring opposite or identical topological charges are shown to exist. In the simplest geometries with two amplifying rings vortex twins with equal topological charges exhibit asymmetric intensity distributions, while vortex anti-twins may be symmetric or asymmetric, depending on the gain level and separation between rings. Different arrangements of amplifying rings allow generation of stable multi-vortex soliton complexes with various topologies, with twins and anti-twins as building blocks.

Vortex solitary-wave structures attract continuously renewed interest because of their unique properties associated to their intensity and phase shapes and to the corresponding topological charges carried by the beams (see, e.g., [1]). In most uniform materials with local nonlinearities vortex solitons, being higher-order excited states, are prone to strong azimuthal modulation instabilities. Such instabilities may be suppressed by several mechanisms, including competing nonlinearities [2,3], transverse modulation of the refractive index in the form of different types of optical lattices [4-6], or nonlocality of the nonlinear response [7]. The simplest vortex solitons carry only one phase dislocation. Nevertheless, complex vortex states that include multiple phase dislocations nested in a single wavefront also exist and they show specific properties. In conservative settings such states, known as multi-vortex solitons, have been studied in triangular [8] and hexagonal [9] optical lattices, as well as in materials with nonlocal nonlinearities [10,11].

Vortex solitary-wave structures may also exist in several dissipative settings [12,13], such as laser amplifiers [14], and systems described by the complex cubic-quintic Ginzburg-Landau equation [15-17], or Bose-Einstein condensates [18,19]. Such states exhibit significant dynamical shape transformations upon their evolution. In particular, it has been recently shown that the evolution of nonlinear excitations in dissipative media is strongly affected by a spatially modulation of the gain profile [20-24]. For example, ring-shaped gain landscapes imprinted in focusing Kerr media with strong two-photon absorption have been shown to support stable stationary [23] and rotating vortex solitons [24].

The aim of this Letter is to show that stationary, twin and anti-twin, multi-vortex soliton complexes can be supported and are stable in suitable gain landscapes made of well-separated or strongly overlapping amplifying rings imprinted in a Kerr nonlinear medium with strong two-photon absorption. In such a setting, the interplay between diffraction and nonlinearity and between localized gain and two-photon absorption results not only in the suppression of the instabilities of the background, but also in complete elimination of both, collapse and destructive azimuthal modulation instabilities characteristic of two-dimensional geometries. We found that vortex twins form as pairs with identical topological charges and they always exhibit asymmetric shapes. Vortex anti-twins carry two opposite charges and can be symmetric or asymmetric, depending on the gain level.

We analyze the evolution of a laser beam propagating in a Kerr nonlinear medium with two-photon absorption and spatially nonuniform gain, that is described by the nonlinear Schrödinger equation for the dimensionless light field amplitude $q$:

$$i\frac{\partial q}{\partial \xi} = -\frac{1}{2}\left(\frac{\partial^2 q}{\partial \eta^2} + \frac{\partial^2 q}{\partial \zeta^2}\right) - q|q|^2 + i\gamma(\eta,\zeta)q - i\alpha q|q|^2. \quad (1)$$

The transverse coordinates $\eta, \zeta$ and the propagation distance $\xi$ are normalized to the characteristic transverse width and to the diffraction length, respectively; the function $\gamma(\eta,\zeta)$ describes the transverse gain distribution; the parameter $\alpha$ characterizes the strength of the two-photon absorption.

We search for stationary solutions in gain landscapes containing several amplifying rings described by the function $\gamma = p_j \sum_k \exp[-(r_k - r_c)^2/d^2]$, where $p_i$ is the gain parameter, $r_k^2 = (\eta - \eta_k)^2 + (\zeta - \zeta_k)^2$, $\eta_k, \zeta_k$ stand for the coordinates of the center of the $k$-th amplifying ring, and $d$ and $r_c$ are the width and radius of each amplifying ring, respectively. The distance between the centers of the neighboring rings is equal to $2sr_c$, i.e. at $s=1$ the maxima of two rings overlap (this results in an increase of the effective amplification in the overlap region), while at $s \to \infty$ the distance between rings becomes infinitely large and one recovers the case of a single amplifying ring. Here we set $r_c = 5.25$, $d = 3.1$, and $\alpha = 2.5$, and use the gain level $p_i$ and separation between rings $s$ as control parameters. However, we verified that the results reported here remain qualitatively similar for other values of $r_c, d$, and $\alpha$. Note that solitons in dissipative nonlinear media exist only if the double balance between gain and losses and between nonlinearity and diffraction is achieved. The field of the stationary solutions can be written in the form $q(\eta,\zeta,\xi) = w(\eta,\zeta)\exp(ib\xi)$, where $w(\eta,\zeta)$ is a function describing the multi-vortex shape and $b$ is the propagation constant. Each individual vortex, that forms in the amplifying ring, is characterized by its topological charge, or winding number, $m$. We consider the simplest case with $m = \pm 1$, although higher-charge vortices may also form complexes. The multi-vortex soliton attractors were obtained by solving Eq. (1) with a standard split-step fast Fourier method up to large distances, i.e., $\xi \sim 10^4$, with input conditions consisting of several rings with suitable wavefront distributions. Usually a 90*45 integration window with 1024*512 transverse points were used, and we verified that results do not depend on such choice. Upon numerical propagation, such inputs emit radiation away, reshape, and quickly converge to stable attractors that remain invariable upon further propagation. The families of solutions were obtained by changing $p_i, s$ and using the output field from a previous step as the input condition for the new set of $p_i, s$ values.

First, we consider the simplest gain landscape with two amplifying rings. Such gain landscapes support vortex twins and anti-twins (Fig. 1). The multi-vortex complexes exist within a wide range of gain parameters and for different separations between the amplifying rings. Within the vortex-antivortex family two types of solitons were found: solutions whose field modulus distributions are symmetric with respect to both $\eta, \zeta = 0$ axes [Fig. 1(a)] and solutions that are asymmetric with respect to vertical axis $\eta = 0$ [Fig. 1(b)]. In all cases vortex twins are asymmetric with respect to both $\eta, \zeta = 0$ axes [Fig. 1(c)]. A close inspection of the phase distribution of the multi-vortex states reveals that in addition to the main phase dislocations residing in the centers of the amplifying rings, secondary dislocations may appear in low-intensity regions. Such secondary vortices ensure a smooth joining of the complex internal currents (that are not circular, unlike in usual radially symmetric vortices) inside multi-vortex states, but at the same time they result in symmetry breaking, especially when dislocations appear not in pairs [Fig. 1(c)]. In the domain between amplifying rings the field modulus features an interference pattern that becomes smoother with increasing separation $s$ between the amplifying rings.

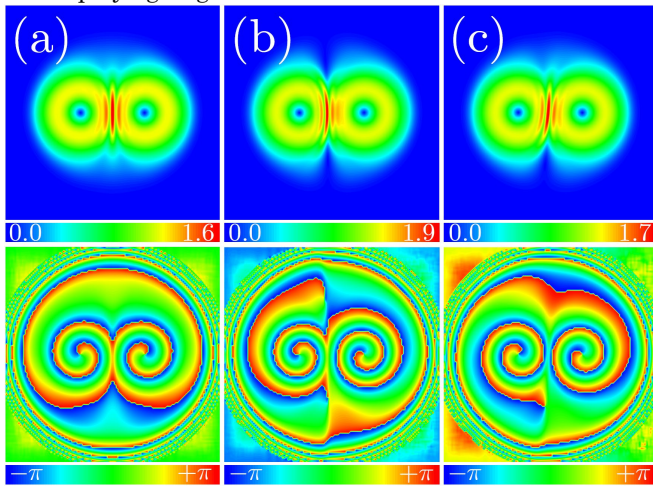

Figure 1. (a) Symmetric vortex-antivortex mode at $p_i = 3.7$. (b) Asymmetric vortex-antivortex soliton at $p_i = 5$. (c) Asymmetric vortex-vortex mode at $p_i = 4$. Top row - field modulus distributions, bottom row - phase distributions. In all cases $s = 1.5$.

By the very nature of dissipative solitons, their energy flow $U$ and propagation constant $b$ are determined by the gain and losses. Figure 2(a) shows the energy flow (defined as $U = \int \int_{-\infty}^{\infty} |q|^2 d\eta d\zeta$) of the vortex anti-twins versus the gain parameter $p_i$. Despite the smooth growth of the energy flow with $p_i$, the symmetry of the solution may change along the curve. Thus, for $s = 1.5$ the soliton mode is symmetric for $1.45 \leq p_i \leq 2.2$ and $2.7 \leq p_i \leq 3.75$ (black branches) and is asymmetric for other $p_i$ values (red branches). The asymmetry becomes more pronounced with increasing gain. The dependence $U(p_i)$ for asymmetric vortex-twins (not shown) is similar to that of vortex anti-twin states. We found that, as in the case of single vortex solitons supported by a single gain ring [23], stable multivortex solitons do not exist below a threshold gain level [the curve in Fig. 2(a) stops at the corresponding $p_i = p_i^{\min}$ value]. For smaller values, the existing absorption cannot be compensated by the available gain. It is worth stressing that only symmetric vortex-antivortex soliton modes exist when $p_i \to p_i^{\min}$. At the same time, no upper threshold in gain was found, although only strongly asymmetric states can be generated at high $p_i$ values.

Of particular interest is the influence of the distance between the centers of the amplifying rings on the energy flow [Fig. 2(b)]. Despite the fact that decreasing the separation results in overlapping amplifying rings and in an effective growth of the gain, we found that both vortex-vortex and vortex-antivortex solitons can exist only above a minimal ring separation. For both types of solitons the energy flow is a nonmonotonic function of the separation. At moderate gain levels, e.g., $p_i = 4$, only a small portion of the vortex-antivortex branch for $1.43 \leq s \leq 1.66$ is asymmetric. Notice that the asymmetric vortex-vortex family terminates exactly for the same $s$ value as the symmetric vortex-antivortex family. Both dependencies $U_{va}$ and $U_{vv}$ approach the same limiting value at $s \to \infty$, which corresponds to the energy of two vortex solitons supported by a single ring.

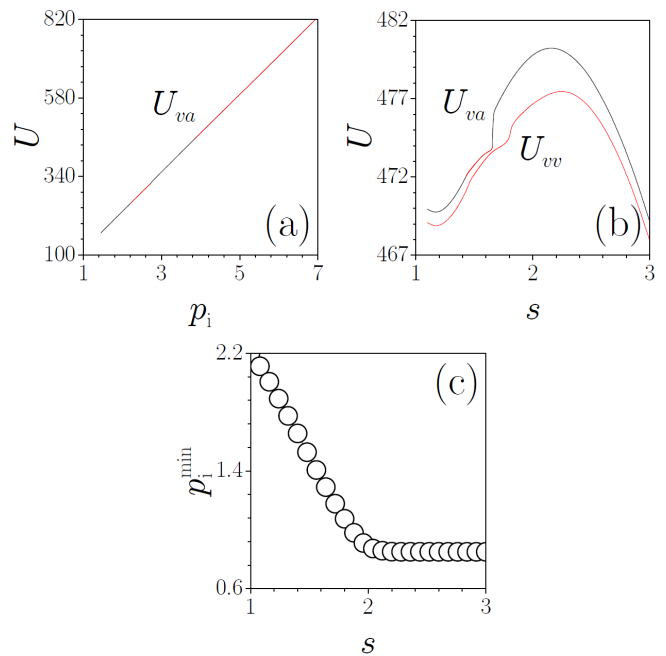

Figure 2. (a) Energy flow of vortex-antivortex soliton versus $p_i$ at $s = 1.5$. (b) Energy flow of vortex-vortex and vortex-antivortex solitons versus $s$ at $p_i = 4$. The branches corresponding to symmetric solitons are shown black, while branches of asymmetric solitons are shown red. (c) Minimal gain required for existence of stable vortex-antivortex solitons versus $s$.

As it was mentioned above, both vortex-vortex and vortex-antivortex families may be stable only above a minimum gain level $p_i^{\min}$. In Fig. 2(c) this threshold gain is plotted as a function of the separation between the amplifying rings. Somehow surprisingly, for weakly separated rings a larger gain is necessary for the stabilization of the multi-vortex complexes. The minimal threshold gain is found to rapidly diminish with increasing separation between rings; it saturates already at $s \sim 2$, when the vortex soliton modes residing in different rings weakly affect each other. Also, we found all vortex complexes studied in the Letter to be rather robust objects. In particular, we verified numerically their stability against the influence of small perturbations.

Besides vortex-pairs, more complex stable multi-vortex states were also found. In particular, we were able to generate a variety of complexes composed of three and four vortex-antivortices. However, it becomes more and more difficult to generate states with symmetric field modulus distributions in structures with larger number of amplifying channels. In contrast, the domains of existence of stable multi-vortex complexes in $(s, p_i)$ plane do not change qualitatively with the increase of number of amplifying rings or with modifications in their mutual arrangement.

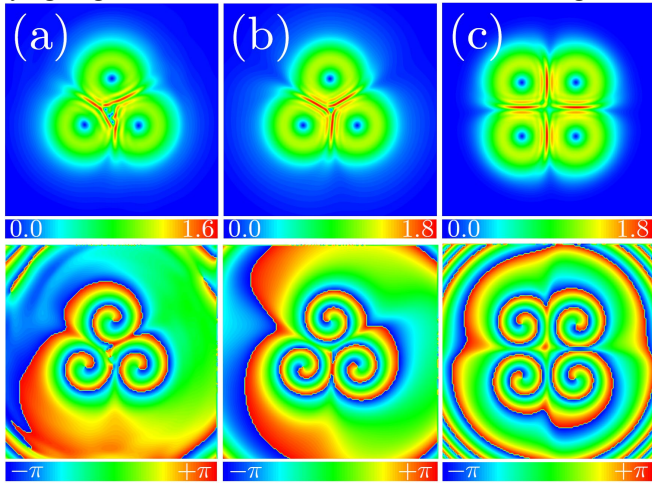

Figure 3. Multivortex solitons in gain landscapes with three (a), (b) and four (c) rings. In (a) $p_i = 3$, while in (b), (c) $p_i = 4$. Top row shows field modulus distributions, while bottom row shows phase distributions. In all cases $s = 1.5$.

Examples of states including three vortices with identical topological charges supported by a triangular system of amplifying rings are shown in Fig. 3(a). One can observe additional secondary multiple dislocations in the region between amplifying rings. For comparison, only one such dislocation is visible in the vortex-vortex complex show in Fig. 1(c). As mentioned above, such additional dislocations are needed to allow closed-contour energy circulation in the wavefront; they thus appear in the stationary vortex for a variety of inputs. The pattern becomes somewhat smoother when one of the vortices carries an opposite topological charge [Fig. 3(b)]. Figure 3(c) shows a stable structure comprising two pairs of vortices and antivortices. Different arrangements of the dislocations in such complexes generate various stable soliton families with slightly different energy flows. All these families may be completely stable for the same set of system parameters although it has to be stressed that they are obtained using different initial conditions.

Summarizing, we showed that several amplifying rings imprinted in a Kerr nonlinear media with two-photon absorption can support rich families of dissipative multi-core, multi-vortex solitons. Such vortex complexes were found to be completely stable. Vortex twins and anti-twins can be symmetric or strongly asymmetric, depending on the arrangement of individual vortices in the neighboring amplifying rings and on the gain strength. Finally, it is worth stressing that the addressed vortex soliton states have no counterparts in conservative or uniform dissipative systems.